\documentclass[aps,prl,twocolumn,showpacs,superscriptaddress]{revtex4}

\usepackage{graphicx}
\usepackage[english]{babel}
\usepackage{bm}    %% For bold math symbols
\def\reff#1{(\ref{#1})}
\newcommand{\be}{\begin{equation}}
\newcommand{\ee}{\end{equation}}
\newcommand{\<}{\langle}
\renewcommand{\>}{\rangle}

\def\spose#1{\hbox to 0pt{#1\hss}}
\def\ltapprox{\mathrel{\spose{\lower 3pt\hbox{$\mathchar"218$}}
 \raise 2.0pt\hbox{$\mathchar"13C$}}}
\def\gtapprox{\mathrel{\spose{\lower 3pt\hbox{$\mathchar"218$}}
 \raise 2.0pt\hbox{$\mathchar"13E$}}}

\newcommand{\scrc}{{\cal C}}

\newcommand{\scrs}{{\cal S}}

\begin{document}

\title{Ferromagnetic phase transition for the spanning-forest model \\
       ($\boldsymbol{q \to 0}$ limit of the Potts model)
       in three or more dimensions}

\author{Youjin Deng}
\email{yd10@nyu.edu}
\affiliation{Department of Physics, New York University,
      4 Washington Place, New York, NY 10003, USA}
\author{Timothy M.~Garoni}
\email{garoni@nyu.edu}
\affiliation{Department of Physics, New York University,
      4 Washington Place, New York, NY 10003, USA}
\author{Alan D. Sokal}
\email{sokal@nyu.edu}
\affiliation{Department of Physics, New York University,
      4 Washington Place, New York, NY 10003, USA}
\affiliation{Department of Mathematics,
      University College London, London WC1E 6BT, UK}

%%\date{\today {\bf\  --- Need to fix final date!!!}}
\date{6 October 2006; published 17 January 2007}

\begin{abstract}
We present Monte Carlo simulations of the spanning-forest model
($q \to 0$ limit of the ferromagnetic Potts model)
in spatial dimensions $d=3,4,5$.
We show that, in contrast to the two-dimensional case,
the model has a ``ferromagnetic'' second-order phase transition
at a finite positive value $w_c$.
We present numerical estimates of $w_c$
and of the thermal and magnetic critical exponents.
We conjecture that the upper critical dimension is 6.
\end{abstract}

\pacs{05.50.+q, 11.10.Kk, 64.60.Cn, 64.60.Fr}

\keywords{Potts model, Fortuin--Kasteleyn representation,
random-cluster model, spanning forests, phase transition, Monte Carlo.}

\maketitle

%%\section*{Introduction}

The Potts model \cite{Potts_52,Wu_82+84}
plays an important role in the modern theory of phase transitions
and critical phenomena,
and is characterized by two parameters:
the number $q$ of Potts spin states,
and the nearest-neighbor coupling $v = e^{\beta J}-1$.
Initially $q$ is a positive integer
and $v$ is a real number in the interval $-1 \le v < +\infty$,
but the Fortuin--Kasteleyn (FK) representation \cite{FK_69+72}
shows that the partition function $Z_G(q,v)$ of the $q$-state Potts model
on any finite graph $G$ is in fact a {\em polynomial}\/ in $q$ and $v$.
This allows us to interpret $q$ and $v$ as taking arbitrary
real or even complex values,
and to study the phase diagram of the Potts model in
the real $(q,v)$-plane or in complex $(q,v)$-space.
In particular, when $q,v > 0$ the FK representation has positive weights
and hence can be interpreted probabilistically
as a correlated bond-percolation model:
the \emph{FK random-cluster model} \cite{Grimmett_06}.
In this way we can study all positive values of $q$,
integer or noninteger, within a unified framework.

In two dimensions, the behavior of the ferromagnetic ($v > 0$)
Potts/random-cluster model is fairly well understood,
thanks to a combination of exact solutions \cite{Baxter_book},
Coulomb-gas methods \cite{Nienhuis_84}
and conformal field theory \cite{DiFrancesco_97}.
But in dimension $d \ge 3$, many important aspects remain unclear:
the location of the crossover between
second-order and first-order behavior \cite{fn_1st2nd};
the nature of the critical exponents and their dependence on $q$;
the value of the upper critical dimension for noninteger $q$;
and the qualitative behavior of the critical curve $v_c(q)$ near $q=0$.

Interesting special cases of the random-cluster model
arise in the limit $q \to 0$.
In particular, the limit $q,v \to 0$ with $w = v/q$ held fixed
gives rise to a model of \emph{spanning forests},
i.e.\ spanning subgraphs without cycles,
in which each occupied edge gets a weight $w$
\cite{forest_refs}.
Very recently, it was shown \cite{CJSSS_prl} ---
generalizing Kirchhoff's matrix-tree theorem \cite{Kirchhoff_et_al} ---
that this spanning-forest model can be mapped onto
a fermionic (Grassmann) theory involving a quadratic (Gaussian) term
and a special nearest-neighbor four-fermion term.
Moreover, this fermionic model possesses an $OSP(1|2)$ supersymmetry
and can be mapped,
to all orders of the perturbation theory in powers of $1/w$,
onto an $N$-vector model [$O(N)$-invariant $\sigma$-model]
analytically continued to $N=-1$.
It follows that, in two dimensions,
the spanning-forest model is perturbatively asymptotically free,
in close analogy to (large classes of) two-dimensional $\sigma$-models
and four-dimensional nonabelian gauge theories.
In particular, the only ferromagnetic ($w>0$) critical point
lies at $w_c = +\infty$,
in agreement \cite{Wu_78} with the exact solutions
on the square, triangular and hexagonal lattices \cite{Baxter_book}
showing that $v_c(q) \propto q^{1/2}$ as $q \downarrow 0$.
%% See also \cite{JSS_forests,Jacobsen-Saleur,CSS_inprep}
%% for more recent work on these connections.

In this Letter we study the spanning-forest model
in spatial dimensions $d \ge 3$, using Monte Carlo methods.
We will show that, in contrast to the two-dimensional case,
the model has a ``ferromagnetic'' second-order phase transition
at a finite positive value $w_c$,
and we will estimate the thermal and magnetic critical exponents
as well as a universal amplitude ratio.
It follows that $v_c(q) \propto q$ as $q \downarrow 0$.
Indeed, we see the present study of the spanning-forest model
as the first step in a comprehensive study of the
random-cluster model as a function of (noninteger)~$q$.

For the random-cluster model with $q \ge 1$,
a collective-mode Monte Carlo algorithm has recently
been invented by Chayes and Machta \cite{Chayes-Machta};
it generalizes the well-known Swendsen--Wang algorithm \cite{Swendsen_87}
and reduces to (a slight variant of) it when $q$ is an integer.
But for $q < 1$, the only available algorithm seems to be
the Sweeny algorithm \cite{Sweeny_83},
which is a local bond-update algorithm.
Ordinarily one would expect such a local algorithm
to exhibit severe critical slowing-down,
at least when the specific heat is divergent \cite{Li-Sokal}.
But the random-cluster model with $q < q_0(d) \approx 2$
has a non-divergent specific heat (i.e., critical exponent $\alpha < 0$),
which suggests that the critical slowing-down might not be so severe after all.
Indeed, our numerical studies of the spanning-forest model
(i.e., the $q \to 0$ limit)
in dimensions $d=2,3,4,5$ strongly suggest
that there is \emph{no} critical slowing-down,
i.e., the dynamic critical exponent $z_{\rm exp}$
associated to the exponential autocorrelation time is zero.
Better yet, the exponent $z_{{\rm int},{\cal O}}$ associated
to the \emph{integrated} autocorrelation time \cite{Sokal_Cargese_96}
turns out to be \emph{negative} for ``global'' observables
such as the mean-square cluster size;
that is, one ``effectively independent'' sample
can be obtained in a time much \emph{less} than a single ``sweep'' ---
a kind of ``critical speeding-up''.

On the other hand, the Sweeny algorithm for $q \neq 1$ requires
a non-local connectivity check each time one tries to update a single bond.
If done in the naive way (e.g., by depth-first or breadth-first search),
this would require a CPU time of order the mean cluster size
$\chi \propto L^{\gamma/\nu} = L^{\approx 2}$
per ``hit'' of a single bond,
leading to a severe ``computational critical slowing-down''.
Recent work by computer scientists on dynamic connectivity algorithms
\cite{dynamic_connectivity}
shows how this can be reduced to $(\log L)^p$,
but at the expense of fairly complicated algorithms and data structures.
We therefore adopted an intermediate solution:
a simple ``homemade'' dynamic connectivity algorithm
that empirically has a slowing-down $L^{\approx 0.7}$.
The details of this algorithm,
along with measurements of the dynamic critical behavior
of the Sweeny algorithm in the spanning-forest limit,
will be reported separately \cite{forests_3d4d_fullpaper}.

We simulated the spanning-forest model in dimensions $d=3,4,5$
on hypercubic lattices of size $L^d$ with periodic boundary conditions.
We measured the cluster-size moments
$\scrs_k = \sum_{{\rm clusters}\:\scrc} \#(\scrc)^k$ for $k=0,2,4$.
We focussed attention on the ratio
$R = \<\scrs_4\> / \<\scrs_2^2\>$,
which tends in the infinite-volume limit to 0 in a disordered phase
and to 1 in an ordered phase,
and is therefore diagnostic of a phase transition.
We also studied $\< \scrs_2 \>$ in order to estimate the
magnetic critical exponent.

In each dimension, we began by making a ``coarse'' set of runs
covering a wide range of $w$ values, using modest-sized lattices
and modest statistics.  If the plots of $R$ versus $w$ indicated
a likely phase transition, we then made a ``fine'' set of runs
covering a small neighborhood of the estimated critical point,
using larger lattices and larger statistics.
Finally, using the results from these latter runs,
we made a ``super-fine'' set of runs extremely close to
the estimated critical point,
using as large lattices and statistics as we could manage,
with the goal of obtaining precise quantitative estimates
of the critical point $w_c$ and the critical exponents.
The complete set of runs reported in this Letter
used approximately 7 years CPU time on a 3.2 GHz Xeon EM64T processor.
% Here we have scaled 1 mafalda year ~~ 0.6 guille years.

\begin{figure}[t]
%\vspace*{0cm} \hspace*{-0cm}
\begin{center}
\includegraphics[width=3.5in]{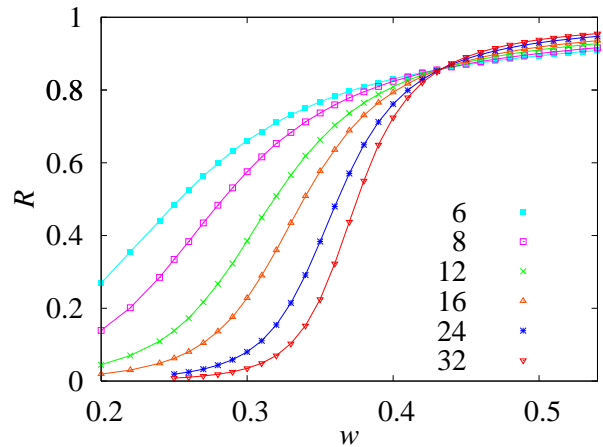}
%\quad \vspace{5cm}  %% TEMPORARY UNTIL FILE IS THERE
\end{center}
\vspace*{-6mm}
\caption{
   Coarse plot of $R$ versus $w$ for spanning forests
   in dimension $d=3$ and lattice sizes $6 \le L \le 32$.
}
\label{fig1}
\end{figure}

\begin{figure}[t]
%\vspace*{0cm} \hspace*{-0cm}
\begin{center}
\includegraphics[width=3.5in]{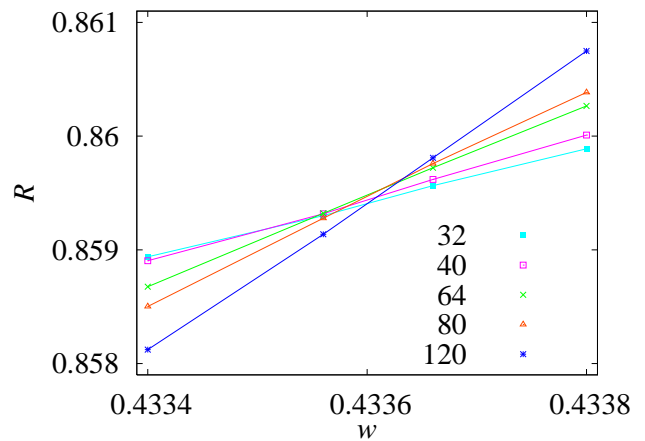}
%\quad \vspace{5cm}  %% TEMPORARY UNTIL FILE IS THERE
\end{center}
\vspace*{-6mm}
\caption{
   ``Super-fine'' plot of $R$ versus $w$ for spanning forests
   in dimension $d=3$ and lattice sizes $32 \le L \le 120$.
}
\label{fig2}
\end{figure}

\begin{figure}[t]
%\vspace*{0cm} \hspace*{-0cm}
\begin{center}
\includegraphics[width=3.5in]{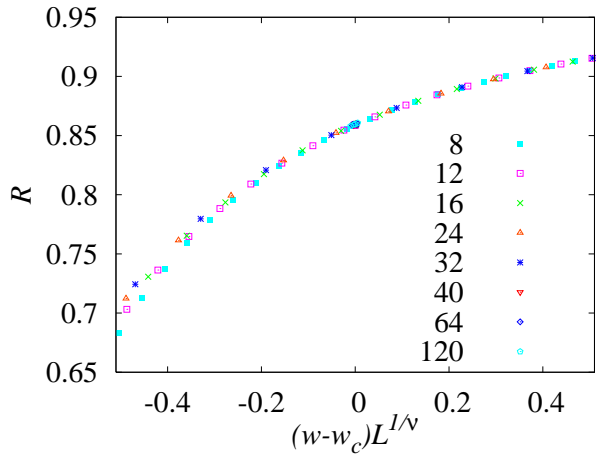}
%\quad \vspace{5cm}  %% TEMPORARY UNTIL FILE IS THERE
\end{center}
\vspace*{-6mm}
\caption{
   Finite-size-scaling plot of $R$ versus $(w-w_c) L^{1/\nu}$,
   with $w_c = 0.43365$ and $\nu = 1.28$,
   for spanning forests in dimension $d=3$ and lattice sizes $8 \le L \le 120$.
}
\label{fig3}
\end{figure}

\begin{figure}[t]
\vspace*{-9mm} \hspace*{-0cm}
\begin{center}
\includegraphics[width=3.5in]{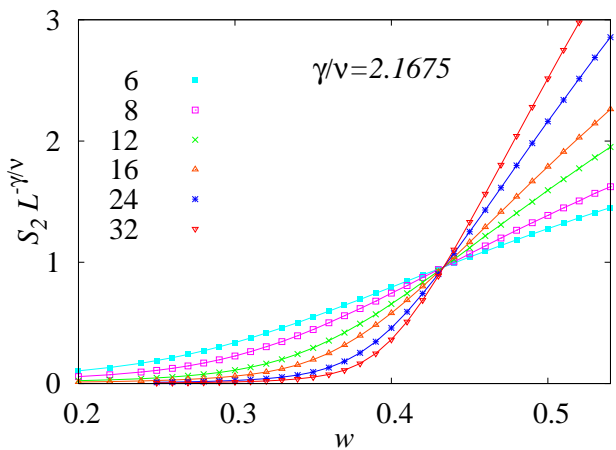}
%\quad \vspace{5cm}  %% TEMPORARY UNTIL FILE IS THERE
\end{center}
\vspace*{-6mm}
\caption{
   Plot of $\<\scrs_2\>/L^{\gamma/\nu}$ versus $w$,
   with $\gamma/\nu = 2.1675$, for spanning forests
   in dimension $d=3$ and lattice sizes $6 \le L \le 32$.
}
\label{fig4}
\end{figure}

The ``coarse'' plot of $R$ versus $w$ for dimension $d=3$
%%% and lattice sizes $L=6,8,12,16,24,32$ is shown in Figure~\ref{fig1},
and lattice sizes $6 \le L \le 32$ is shown in Figure~\ref{fig1},
and shows a clear order-disorder transition at $w_c \approx 0.43$.
The corresponding ``super-fine'' plot,
for lattice sizes $32 \le L \le 120$, is shown in Figure~\ref{fig2}.
We fit the data to Ans\"atze obtained from
\begin{eqnarray}
   & &
   R \;=\;  R_c \,+\, a_1 (w-w_c) L^{1/\nu} \,+\, a_2 (w-w_c)^2 L^{2/\nu}
        \nonumber \\
   & & \qquad\qquad
     \:+\, b_1 L^{-\omega_1} \,+\, b_2 L^{-\omega_2} \,+\, \ldots
 \label{eq.R_Ansatz}
\end{eqnarray}
by omitting various subsets of terms,
and we systematically varied $L_{\rm min}$
(the smallest $L$ value included in the fit).
We also made analogous fits for $\< \scrs_2 \> / L^{\gamma/\nu}$.
Comparing all these fits,
we estimate the critical point $w_c = 0.43365 \pm 0.00002$,
the critical exponents $\nu = 1.28 \pm 0.04$
and $\gamma/\nu = 2.1675 \pm 0.0010$,
and the universal amplitude ratio $R_c = 0.8598 \pm 0.0003$
(68\% subjective confidence intervals, including both statistical error
 and estimated systematic error due to unincluded corrections to scaling).
% {\bf How to estimate systematic error on $\gamma/\nu$ and $R_c$???
%    Maybe do fits with $w_c$ and $y_t$ \emph{imposed} at a variety of
%    plausible values, and see how much $\gamma/\nu$ and $R_c$ move???}
A finite-size-scaling plot using these parameters
is shown in Figure~\ref{fig3}.
A ``coarse'' plot of $\< \scrs_2 \> / L^{\gamma/\nu}$
using the estimated value of $\gamma/\nu$
is shown in Figure~\ref{fig4}.

%%%%%%%%%%%%%%%%%%%%%%%%%%%%%%%%%%%%%%%%%%%%%%%%%%%%%%%%%%%%%%%%%%%%%%%%%%%%%%
%
% FIGURE 1
%

%%%%%%%%%%%%%%%%%%%%%%%%%%%%%%%%%%%%%%%%%%%%%%%%%%%%%%%%%%%%%%%%%%%%%%%%%%%%%%
%
% FIGURE 2
%

%%%%%%%%%%%%%%%%%%%%%%%%%%%%%%%%%%%%%%%%%%%%%%%%%%%%%%%%%%%%%%%%%%%%%%%%%%%%%%
%
% FIGURE 3
%

%%%%%%%%%%%%%%%%%%%%%%%%%%%%%%%%%%%%%%%%%%%%%%%%%%%%%%%%%%%%%%%%%%%%%%%%%%%%%%
%
% FIGURE 4
%

%%%%%%%%%%%%%%%%%%%%%%%%%%%%%%%%%%%%%%%%%%%%%%%%%%%%%%%%%%%%%%%%%%%%%%%%%%%%%%

\begin{figure}[t]
%\vspace*{0cm} \hspace*{-0cm}
\begin{center}
\includegraphics[width=3.5in]{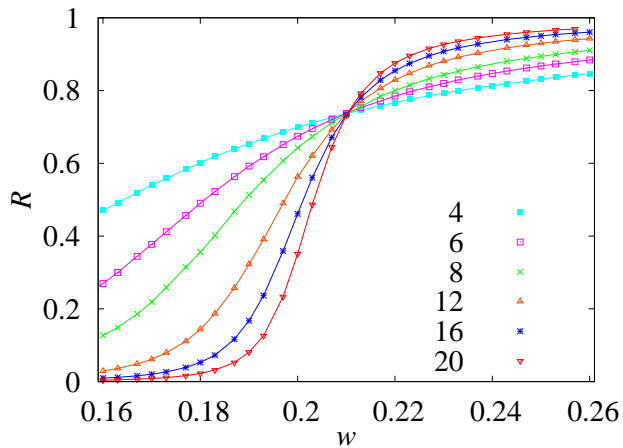}
%\quad \vspace{5cm}  %% TEMPORARY UNTIL FILE IS THERE
\end{center}
\vspace*{-6mm}
\caption{
   Coarse plot of $R$ versus $w$ for spanning forests
   in dimension $d=4$ and lattice sizes $4\le L \le 20$.
}
\label{fig5}
\end{figure}

\begin{figure}[t]
%\vspace*{0cm} \hspace*{-0cm}
\begin{center}
\includegraphics[width=3.5in]{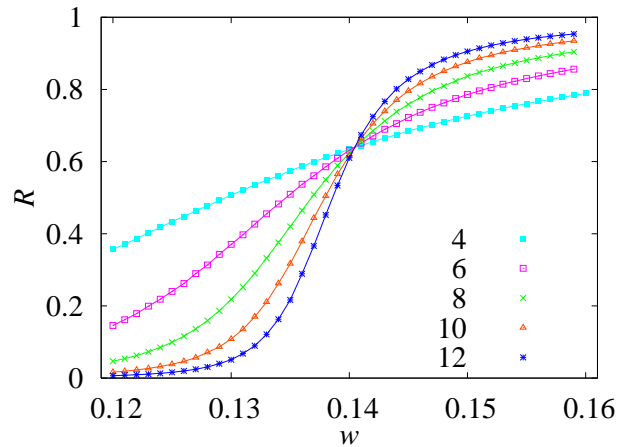}
%\quad \vspace{5cm}  %% TEMPORARY UNTIL FILE IS THERE
\end{center}
\vspace*{-6mm}
\caption{
   Coarse plot of $R$ versus $w$ for spanning forests
   in dimension $d=5$ and lattice sizes $4\le L \le 12$.
}
\label{fig6}
\end{figure}

The ``coarse'' plots of $R$ versus $w$ for dimensions $d=4,5$
are shown in Figures~\ref{fig5} and \ref{fig6}, respectively.
Once again they show a clear order-disorder transition.
For lack of space, we refrain from showing the corresponding
``super-fine'' plots
(which use lattice sizes up to $64^4$ and $20^5$)
and simply give the results of fits
to Ans\"atze of the general type \reff{eq.R_Ansatz}.
In dimension $d=4$, we estimate
$w_c = 0.210302 \pm 0.000010$,
$\nu = 0.80 \pm 0.01$,
$\gamma/\nu = 2.1603 \pm 0.0010$ and
$R_c = 0.73907 \pm 0.00010$.
In dimension $d=5$, we estimate
$w_c = 0.14036 \pm 0.00002$,
$\nu = 0.59 \pm 0.02$,
$\gamma/\nu = 2.08 \pm 0.02$ and
$R_c = 0.625 \pm 0.015$.

%%%%%%%%%%%%%%%%%%%%%%%%%%%%%%%%%%%%%%%%%%%%%%%%%%%%%%%%%%%%%%%%%%%%%%%%%%%%%%
%
% FIGURE 5
%

%%%%%%%%%%%%%%%%%%%%%%%%%%%%%%%%%%%%%%%%%%%%%%%%%%%%%%%%%%%%%%%%%%%%%%%%%%%%%%
%
% FIGURE 6
%

%%%%%%%%%%%%%%%%%%%%%%%%%%%%%%%%%%%%%%%%%%%%%%%%%%%%%%%%%%%%%%%%%%%%%%%%%%%%%%

In Table~\ref{table1} we summarize the estimated critical exponents
for ferromagnetic Potts models with $q=0$ (this work),
1 (percolation) and 2 (Ising) in dimensions $d=2,3,4,5$.
It is evident that $\nu$ varies quite sharply as a function of $q$ and $d$,
while $\gamma/\nu$ varies much more slowly.
The dimension-dependences of $\nu$ and $\gamma/\nu$ for $q=0$
are consistent with the conjecture that
they are tending to the mean-field values $1/2$ and $2$ in dimension $d=6$,
just as they do for $q=1$.
This in turn supports the more general conjecture
that the upper critical dimension is 6
for all random-cluster models with $0 \le q < 2$,
and is 4 only when $q=2$.

%%%%%%%%%%%%%%%%%%%%%%%%%%%%%%%%%%%%%%%%%%%%%%%%%%%%%%%%%%%%%%%%%%%%%%%%%%%%%%
%
% TABLE 1
%

\begin{table}[t]
\vspace*{5mm} \hspace*{-0cm} %% PUSH FLOAT DOWN A BIT
\begin{center}
\begin{tabular}{|c||@{}c@{}|@{}c@{}|@{}c@{}|}
\cline{2-4}
\multicolumn{1}{c|}{\quad}    & $q=0$ & $q=1$ & $q=2$  \vphantom{\Big(} \\
\cline{2-4} \multicolumn{4}{c}{\quad} \\[-3mm] \hline
$d=2$ &  $\begin{array}{c}  \nu=\infty \\ \gamma/\nu=2  \end{array}$
      &  $\begin{array}{c}  \nu=4/3    \\ \gamma/\nu=43/24  \end{array}$
      &  $\begin{array}{c}  \nu=1      \\ \gamma/\nu=7/4  \end{array}$
  \\
\hline
$d=3$ &  $\begin{array}{c}  \nu=1.28(4) \\ \gamma/\nu=2.1675(10)  \end{array}$
      &  $\begin{array}{c}  \nu=0.874(2) \\ \gamma/\nu=2.0455(6)  \end{array}$
      &  $\begin{array}{c}  \nu=0.6301(5)  \\ \gamma/\nu=1.9634(5) \end{array}$
  \\
\hline
$d=4$ &  $\begin{array}{c}  \nu=0.80(1) \\ \gamma/\nu=2.1603(10)  \end{array}$
      &  $\begin{array}{c}  \nu=0.689(10)   \\ \gamma/\nu=2.094(3) \end{array}$
      &  $\begin{array}{c}  \nu=1/2\,(\log) \\ \gamma/\nu=2\,(\log) \end{array}$
  \\
\hline
$d=5$ &  $\begin{array}{c}  \nu=0.59(2)    \\ \gamma/\nu=2.08(2)  \end{array}$
      &  $\begin{array}{c}  \nu=0.57(1)    \\ \gamma/\nu=2.08(2)  \end{array}$
      &  $\begin{array}{c}  \nu=1/2     \\ \gamma/\nu=2     \end{array}$
  \\
\hline
\end{tabular}
\end{center}
\vspace*{-3mm}
\caption{
   Critical exponents $\nu$ and $\gamma/\nu$ versus $q$ and $d$.
   $d=2$: presumed exact values \cite{Salas-Sokal_potts4}.
   $d=3,4,5$, $q=0$: this work.
   $d=3$, $q=1$: \cite{d=3_percolation}.
   $d=3$, $q=2$: \cite{d=3_Ising}.
      %% \cite{Ballesteros_99,Hasenbusch_various,Campostrini_02}.
   $d=4$, $q=1$: \cite{Ballesteros_97,Paul_01}.
   $d=5$, $q=1$: \cite{Paul_01,Adler_90}.
   $d=4,5$, $q=2$: presumed exact values.
}
\label{table1}
\end{table}

%%%%%%%%%%%%%%%%%%%%%%%%%%%%%%%%%%%%%%%%%%%%%%%%%%%%%%%%%%%%%%%%%%%%%%%%%%%%%%

This conjecture is supported by a
field-theoretic renormalization-group calculation
in dimension $d=6-\epsilon$ through order $\epsilon^3$ \cite{Alcantara_81}
in which $q=2$ plays a distinguished role
(all the correction terms vanish there).
%
% NEED TO DOUBLE-CHECK THESE EQUATIONS!!!
%
% \begin{eqnarray}
%    \gamma/\nu & = &
%    2 \,+\, {2-q \over 3(10-3q)} \epsilon
%      \,+\, {(2-q) (420 - 257q + 43q^2) \over 27 (10-3q)^3} \epsilon^2
%            \nonumber \\
%    & &
%      \,+\, {(2-q) \over 972 (10-3q)^5}
%            \Bigl[(774720 - 958836q + 464504q^2 - 101525q^3 + 8375q^4)
%                   - 5184 (3-q)(10-3q)(13-6q+q^2) \zeta(3)
%            \Bigr] \epsilon^3
%      \,+\, O(\epsilon^4)
%          \\[2mm]
%    1/\nu  & = &
%    2 \,-\, {5(2-q) \over 3(10-3q)} \epsilon
%      \,-\, {(2-q) (900 - 290q + 43q^2) \over 54 (10-3q)^3} \epsilon^2
%            \nonumber \\
%    & &
%      \,-\, {(2-q) \over 1944 (10-3q)^5}
%            \Bigl[(622800 - 537720q + 330344q^2 - 91790q^3 + 8375q^4)
%                   - 2592 (10-3q)(120-32q+3q^2-2q^3) \zeta(3)
%            \Bigr] \epsilon^3
%      \,+\, O(\epsilon^4)
% \end{eqnarray}
%
Specializing to $q=0$, we have
\begin{eqnarray*}
   \gamma/\nu
   & \!=\! &
   2 + {\epsilon \over 15} + {7 \epsilon^2 \over 225}
     - \left( {26\,\zeta(3) \over 625} - {269 \over 16875} \right) \epsilon^3
     + O(\epsilon^4)
        \\
   & \!=\! &
   2 + 0.066667 \epsilon + 0.031111 \epsilon^2 - 0.034065 \epsilon^3
     + O(\epsilon^4)
\end{eqnarray*}
\begin{eqnarray*}
   1/\nu
   & \!=\! &
   2 - {\epsilon \over 3} - {\epsilon^2 \over 30}
     + \left( {4\,\zeta(3) \over 125} - {173 \over 27000} \right) \epsilon^3
     \,+\, O(\epsilon^4)
        \\
   & \!=\! &
   2 - 0.333333 \epsilon - 0.033333 \epsilon^2 + 0.032058 \epsilon^3
     + O(\epsilon^4)
\end{eqnarray*}
These series seem rather difficult to resum,
especially when $\epsilon \gtapprox 2$,
but they are in qualitative agreement with
the exponents listed in Table~\ref{table1}.
Moreover, a slightly better agreement can be obtained
by imposing the known exact values at $\epsilon=4$
on the interpolating function.

Details of these simulations and their data analysis,
including analysis of universal amplitude ratios other than $R$,
will be reported separately \cite{forests_3d4d_fullpaper}.

\begin{acknowledgments}
We thank Henk Bl\"ote, Sergio Caracciolo, Andrea Pelissetto,
Jes\'us Salas and Andrea Sportiello for helpful discussions.
This work was supported in part by NSF grants PHY--0116590 and PHY--0424082.
%% PHY--0116590 is the MRI computer grant
\end{acknowledgments}

%%%%%%%%%%%%%%%%%%%%%%%%%%%%%%%%%%%%%%%%%%%%%%%%%%%%%%%%%%%%%%%%%%%%%%


\begin{thebibliography}{99}

\bibitem{Potts_52}  R.B. Potts,
   Proc. Cambridge Philos. Soc. {\bf 48}, 106 (1952).

\bibitem{Wu_82+84}  F.Y. Wu, Rev. Mod. Phys. {\bf 54}, 235 (1982);
   {\bf 55}, 315 (E) (1983);
F.Y. Wu, J. Appl. Phys. {\bf 55}, 2421 (1984).

\bibitem{FK_69+72}  P.W. Kasteleyn and C.M. Fortuin,
   J. Phys. Soc. Japan {\bf 26} (Suppl.), 11 (1969);
C.M. Fortuin and P.W. Kasteleyn, Physica {\bf 57}, 536 (1972).

\bibitem{Grimmett_06}  G. Grimmett, {\em The Random-Cluster Model}\/
   (Springer-Verlag, New York, 2006).

\bibitem{Baxter_book} R.J. Baxter, {\em Exactly Solved Models in Statistical
        Mechanics}\/ (Academic Press, London--New York, 1982).

\bibitem{Nienhuis_84}  B. Nienhuis, J. Stat. Phys. {\bf 34}, 731 (1984).

\bibitem{DiFrancesco_97}  P. Di Francesco, P. Mathieu and D. S\'en\'echal,
   {\em Conformal Field Theory}\/ (Springer-Verlag, New York, 1997).

\bibitem{fn_1st2nd}  We stress that this is a lattice-dependent question;
the answer is \emph{not}\/ universal among lattices of a given dimension $d$.
However, for each spatial dimension $d$ one can also ask
what is the maximum $q$ for which there \emph{exists}\/
a local Hamiltonian yielding a second-order transition:
let us call this $q_\star(d)$.
It is believed on the basis of conformal field theory that $q_\star(2) = 4$;
and it is generally expected that $q_\star(d) = 2$ for $d \ge 4$
[A. Aharony and E. Pytte, Phys. Rev. B {\bf 23}, 362 (1981);
 K.E. Newman, E.K. Riedel and S. Muto, Phys. Rev. B {\bf 29}, 302 (1984);
 {\bf 30}, 2924 (E) (1984)].
Little is known about the value of $q_\star(3)$.
See H. Bl\"ote, Y. Deng, X. Qian and A.D. Sokal, in preparation,
for further discussion.

\bibitem{forest_refs}  M.J. Stephen, Phys. Lett. A {\bf 56}, 149 (1976);
F.Y. Wu, J. Phys. A {\bf 10}, L113 (1977);
J.L. Jacobsen, J. Salas and A.D. Sokal,
   J. Stat. Phys. {\bf 119}, 1153 (2005), cond-mat/0401026;
A.D. Sokal,
%% The multivariate Tutte polynomial
%% (alias Potts model) for graphs and matroids,
in {\em Surveys in Combinatorics, 2005}\/,
ed.~B.S. Webb (Cambridge University Press, 2005),
math.CO/0503607.

\bibitem{CJSSS_prl} S. Caracciolo, J.L. Jacobsen, H. Saleur, A.D. Sokal
  and A. Sportiello, Phys. Rev. Lett. {\bf 93}, 080601 (2004),
  cond-mat/0403271.
  See also J.L. Jacobsen and H. Saleur,
     Nucl. Phys. B {\bf 716}, 439 (2005), cond-mat/0502052;
  S. Caracciolo, A.D. Sokal and A. Sportiello,
    in preparation.


\bibitem{Kirchhoff_et_al}
G. Kirchhoff,
   %% \"Uber die Aufl\"osung der Gleichungen, auf welche man bei der
   %% Untersuchung der linearen Verteilung galvanischer Str\"ome
   %% gef\"urht wird,
    Ann. Phys. Chem. {\bf 72}, 497 (1847);
R.L. Brooks, C.A.B. Smith, A.H. Stone and W.T. Tutte,
   %% The dissection of rectangles into squares,
   Duke Math. J. {\bf 7}, 312 (1940);
A. Nerode and H. Shank,
   %% An algebraic proof of Kirchhoff's network theorem,
   Amer. Math. Monthly {\bf 68}, 244 (1961);
S. Chaiken,
   %% A combinatorial proof of the all minors matrix tree theorem,
   SIAM J. Alg. Disc. Meth. {\bf 3}, 319 (1982);
J.W. Moon,
   %% Some determinant expansions and the matrix-tree theorem,
   Discrete Math. {\bf 124}, 163 (1994);
A. Abdesselam,
   %% Grassmann--Berezin calculus and theorems of the matrix-tree type,
   Adv. Appl. Math. {\bf 33}, 51 (2004), math.CO/0306396.
%% {\bf If tight for space, say ``See Abdesselam \ldots and references
%%    cited therein.''}


% \bibitem{Jacobsen-Saleur}  J.L. Jacobsen and H. Saleur,
%    Nucl. Phys. B {\bf 716}, 439 (2005), cond-mat/0502052.
% 
% \bibitem{CSS_inprep}  S. Caracciolo, A.D. Sokal and A. Sportiello,
%   in preparation.
%   %% Both the "hyperforests" and "forests_ON" papers.
%   %%   Put preprint citations when available.


\bibitem{Wu_78}  F.Y. Wu, Phys. Rev. B {\bf 18}, 516 (1978).
   
\bibitem{Chayes-Machta} L. Chayes and J. Machta,
   Physica A {\bf 254}, 477 (1998).
   See also Y. Deng, T.M. Garoni, J. Machta and A.D. Sokal, in preparation.

\bibitem{Swendsen_87} R.H. Swendsen and J.-S. Wang,
   Phys. Rev. Lett. {\bf 58}, 86 (1987).
   See also R.G. Edwards and A.D. Sokal,
   Phys. Rev. D {\bf 38}, 2009 (1988).

\bibitem{Sweeny_83}  M. Sweeny, Phys. Rev. B {\bf 27}, 4445 (1983).
   For variants of the Sweeny algorithm, see
   F. Gliozzi, Phys. Rev. E {\bf 66}, 016115 (2002), cond-mat/0201285
   as corrected by
   J.-S. Wang, O. Kozan and R.H. Swendsen,
   Phys. Rev. E {\bf 66}, 057101 (2002), cond-mat/0206437.

\bibitem{Li-Sokal}  X.-J. Li and A.D. Sokal,
   Phys. Rev. Lett. {\bf 63}, 827 (1989), see last paragraph.

\bibitem{Sokal_Cargese_96}
A.D. Sokal, in {\em Functional Integration: Basics and Applications}\/,
   ed. C. de Witt-Morette, P. Cartier and A. Folacci
   (Plenum, New York, 1997).

\bibitem{dynamic_connectivity}
   D. Alberts, G. Cattaneo and G.F. Italiano,
      ACM J. Exp. Algorithmics  {\bf 2}, no.~5 (1997);
   M.R. Henzinger and V. King, J. ACM {\bf 46}, 502 (1999);
   R.D. Iyer Jr. {\em et al.}\/,
      ACM J. Exp. Algorithmics  {\bf 6}, no.~4 (2001);
   J. Holm, K. de Lichtenberg and M. Thorup, J. ACM {\bf 48}, 723 (2001).

\bibitem{forests_3d4d_fullpaper}  Y. Deng, T.M. Garoni and A.D. Sokal,
   in preparation.

\bibitem{Salas-Sokal_potts4}  See e.g. J. Salas and A.D. Sokal,
   J. Stat. Phys. {\bf 88}, 567 (1997), hep-lat/9607030,
   Appendix A.1 and references cited therein.

\bibitem{d=3_percolation}  H.G. Ballesteros {\em et al.}\/,
   J. Phys. A {\bf 32}, 1 (1999), cond-mat/9805125;
   Y. Deng and H.W.J. Bl\"ote, Phys. Rev. E {\bf 72}, 016126 (2005).

% \bibitem{Hasenbusch_various}  M. Hasenbusch, K. Pinn and S. Vinti,
%   Phys. Rev. B {\bf 59}, 11471 (1999), hep-lat/9806012;
%   M. Hasenbusch, J. Phys. A {\bf 32}, 4851 (1999), hep-lat/9902026;
%   M. Hasenbusch, Int. J. Mod. Phys. C {\bf 12}, 911 (2001).
% 
% \bibitem{Campostrini_02}  M. Campostrini {\em et al.}\/,
%    Phys. Rev. E {\bf 65}, 066127 (2002), cond-mat/0201180.

\bibitem{d=3_Ising}  A. Pelissetto and E. Vicari,
   Phys. Rep. {\bf 368}, 549 (2002), cond-mat/0012164;
   Y. Deng and H.W.J. Bl\"ote, Phys. Rev. E {\bf 68}, 036125 (2003).

\bibitem{Ballesteros_97}  H.G. Ballesteros {\em et al.}\/,
   Phys. Lett. B {\bf 400}, 346 (1997), hep-lat/9612024.

\bibitem{Paul_01} G. Paul, R.M. Ziff and H.E. Stanley,
   Phys. Rev. E {\bf 64}, 026115 (2001), cond-mat/0101136.

\bibitem{Adler_90}  J. Adler {\em et al.}\/, Phys. Rev. B {\bf 41},
   9183 (1990).

\bibitem{Alcantara_81}   O.F. de Alcantara Bonfim, J.E. Kirkham and
   A.J. Mc\-Kane, J. Phys. A {\bf 14}, 2391 (1981).



\end{thebibliography}
\end{document}